# Real-Space Analysis of Two-Photon Polarization States in Type-II SPDC for High-Purity Polarization Entanglement

KOJI TANINAKA[1], TAKASHI KAKUE[1], AND KEN MORITA[1*]
[1]*Graduate School of Engineering, Chiba University, Chiba 263-8522, Japan*
[*]morita@chiba-u.jp

## Abstract

Spontaneous parametric down-conversion (SPDC) is one of the most widely used sources of polarization-entangled photon pairs, and understanding the generated biphoton state is essential for realizing high-brightness and high-purity entangled-photon sources. In particular, Type-II SPDC produces a biphoton wavefunction with a complex coupling between the spatial and polarization degrees of freedom owing to birefringence. In this study, we calculate the real-space distribution of the two-photon polarization state from the biphoton wavefunction of Type-II SPDC generated in a β-barium borate (BBO) crystal. By identifying the spatial correlation direction between polarization and real-space coordinates, we designed aperture shapes that restrict the collection along this correlation direction. Through both numerical simulations and experiments, we demonstrate that such correlation-aligned apertures simultaneously achieve higher entanglement purity and improved photon collection efficiency. These results establish a practical design principle for optimizing aperture geometries based on the real-space biphoton wavefunction, providing a new approach to realizing high-purity and high-brightness SPDC entangled-photon sources.

## 1. Introduction

Spontaneous parametric down-conversion (SPDC) is a representative nonlinear optical process for generating entangled photon pairs [1]. Photon pairs generated through SPDC exhibit correlations in multiple degrees of freedom, including polarization, spatial mode, and frequency [2–6]. Owing to these properties, they have been widely utilized as essential quantum resources for quantum information technologies such as quantum teleportation [7–10], remote state preparation [11,12], quantum communication [13,14], and quantum sensing [15].

For high-performance quantum information applications, it is essential to efficiently collect photon pairs while maintaining a high degree of entanglement purity. In Type-II SPDC, the directional dependence of the phase-matching condition introduces correlations between the real-space and polarization degrees of freedom of the generated photon pairs. As a result, the polarization-entanglement purity strongly depends on the spatial selection imposed by the collection aperture [16–19]. In general, a higher entanglement purity can be achieved by restricting the spatial acceptance of the aperture; however, this inevitably reduces the effective photon-pair collection rate, leading to a trade-off between entanglement purity and brightness. The effect of aperture geometry on entanglement purity has been discussed previously [18]. More recently, full-field measurements have revealed the spatial variation of polarization entanglement across the SPDC field [20]. However, the relationship between this spatial variation and the underlying real-space biphoton state, and its implications for aperture design, remain to be fully understood.

Previous studies on the real-space biphoton state produced by SPDC have often employed several approximations to simplify the analytical treatment of the biphoton wavefunction [21–26]. These include replacing the sinc phase-matching function with a Gaussian function, introducing the sum and difference coordinates, and assuming a slowly varying phase-matching condition around the pump propagation direction. While these approximations are useful for describing the spatial intensity distribution, they cannot accurately describe the complex phase-matching characteristics arising from the anisotropy of nonlinear crystals in Type-II SPDC. In particular, they make it difficult to accurately describe the amplitude and phase structure of the real-space biphoton wavefunction, which plays an important role in achieving both high entanglement purity and high photon-pair collection efficiency.

In this study, we calculate the real-space biphoton wavefunction in Type-II SPDC using a β-barium borate (BBO) crystal by Fourier transforming the momentum-space biphoton wavefunction derived from the interaction Hamiltonian [26,27]. This approach enables us to describe biphoton states while incorporating the complex phase-matching characteristics and phase information arising from crystal anisotropy. As a result, we can accurately describe the real-space distribution of the two-photon polarization state through the coherent superposition of the two Type-II SPDC emission processes. We then analyze how spatial selection by an aperture influences polarization entanglement and investigate the dependence of the entanglement purity on the aperture geometry. Based on these theoretical predictions, we experimentally measure the relationship between brightness and the Bell parameter (CHSH parameter $S$) [28] for different aperture shapes to verify the predicted aperture dependence. The results demonstrate that restricting the collection along the direction exhibiting a strong correlation between polarization and real-space coordinates enables the generation of photon pairs with a higher parameter while maintaining comparable brightness. These findings provide a practical design guideline for achieving both high entanglement purity and high photon-pair collection efficiency.

## 2. Theory

In this section, we theoretically analyze the photon-pair generation process in Type-II SPDC using a BBO crystal and calculate the real-space biphoton wavefunction. In particular, we investigate the spatial distributions of the relative amplitude and phase between the two SPDC emission processes, which characterize the generated polarization-entangled photon pairs. We then analyze how these spatial structures determine the dependence of the polarization-entanglement purity on the aperture shape.

In Type-II SPDC using a BBO crystal, signal and idler photons are generated simultaneously with correlations in both the time–energy and position–momentum degrees of freedom. As illustrated in Fig. 1(a), the generated signal and idler photons are emitted into distinct conical distributions around the pump axis as a result of the phase-matching condition, forming two displaced rings on an observation screen [29,30]. Polarization-entangled states are generated in the intersection region of these two cones. To analyze the spatial properties of the polarization-entangled state formed in the overlap region of the two emission cones, we begin with the biphoton state in momentum space derived from the interaction Hamiltonian within first-order perturbation theory [26,27,31],

$$\left|\psi_q(\boldsymbol{q}_s,\boldsymbol{q}_i)\right\rangle = A\iint d\boldsymbol{q}_s d\boldsymbol{q}_i\, V(\boldsymbol{q}_p)\Phi(\boldsymbol{q}_s,\boldsymbol{q}_i)|\boldsymbol{q}_s\rangle|\boldsymbol{q}_i\rangle \quad (1)$$

where $A$ is a constant independent of momentum, and $V(\boldsymbol{q}_p)$ denotes the angular spectrum of the pump beam. Through the transverse momentum conservation relation ($\boldsymbol{q}_p = \boldsymbol{q}_s + \boldsymbol{q}_i$), $V(\boldsymbol{q}_p)$ determines the spatial correlation between the signal and idler photons. The phase-matching function $\Phi(\boldsymbol{q}_s,\boldsymbol{q}_i) = \exp(i\Delta kL/2)\mathrm{sinc}(\Delta kL/2)$ describes the phase-matching condition for a crystal of length $L$, including the directional dependence and phase information arising from birefringence. Here, $\Delta k$ represents the longitudinal phase mismatch. The quantity $\psi_q(\boldsymbol{q}_s,\boldsymbol{q}_i) = V(\boldsymbol{q}_s+\boldsymbol{q}_i)\Phi(\boldsymbol{q}_s,\boldsymbol{q}_i)$ is the momentum-space biphoton wavefunction.

The positive-frequency field operator in real space is related to the momentum-space field operator through a Fourier transform [26,27] and described as $\hat{E}^{(+)}(\boldsymbol{\rho},z) = \int d\boldsymbol{q}e^{i(k_z z-\boldsymbol{q}\cdot\boldsymbol{\rho})}\hat{a}(\boldsymbol{q})$. Therefore, the real-space biphoton wavefunction can be obtained by applying the inverse Fourier transform to the momentum-space biphoton wavefunction together with the propagation phase. Here, $k_z = \sqrt{(\omega/c)^2-|\boldsymbol{q}|^2}$ is the longitudinal wave-vector component and introduces phase information that depends on the transverse wave vector ($\boldsymbol{q}$). The biphoton state at the transverse positions ($\boldsymbol{\rho}_s,\boldsymbol{\rho}_i$) is then given by

$$\left|\psi_\rho(\boldsymbol{\rho}_s,\boldsymbol{\rho}_i)\right\rangle = \psi_\rho(\boldsymbol{\rho}_s,\boldsymbol{\rho}_i)|\boldsymbol{\rho}_s\rangle|\boldsymbol{\rho}_i\rangle = \mathcal{F}^{-1}\left[\psi_q(\boldsymbol{q}_s,\boldsymbol{q}_i)e^{i(k_{sz}+k_{iz})z}\right]|\boldsymbol{\rho}_s\rangle|\boldsymbol{\rho}_i\rangle. \quad (2)$$

The real-space biphoton wavefunction $\psi_\rho(\boldsymbol{\rho}_s,\boldsymbol{\rho}_i)$ can be interpreted as a two-photon correlation function representing the complex probability amplitude of finding an idler photon at position $\boldsymbol{\rho}_i$ when a signal photon is detected at position $\boldsymbol{\rho}_s$.

Figure 1(b) shows the calculated intensity distributions of the signal and idler photons in momentum-space obtained from the biphoton wavefunction $\psi_q(\boldsymbol{q}_s,\boldsymbol{q}_i)$, while Fig. 1(c) shows the corresponding intensity distributions in real-space calculated from the real-space biphoton wavefunction $\psi_\rho(\boldsymbol{\rho}_s,\boldsymbol{\rho}_i)$. To facilitate comparison with the experiment described later, the calculations were performed using a crystal optic-axis angle of $\theta = 41.8°$, a pump wavelength of $\lambda_p = 405$ nm, signal and idler wavelengths of $\lambda_{s,i} = 810$ nm, and a crystal length of $L = 2$ mm. The pump beam was assumed to have a Gaussian angular spectrum, $V(\boldsymbol{q}_p) = \exp(-|\boldsymbol{q}_p|^2 w_0^2/4)$ with a beam waist of $w_0 = 0.4$ mm. The distance between the crystal output surface and the observation plane was set to 10 cm. It can be seen that the results in Figs. 1(b) and 1(c) correspond to each other through Fourier transformation and free-space propagation.

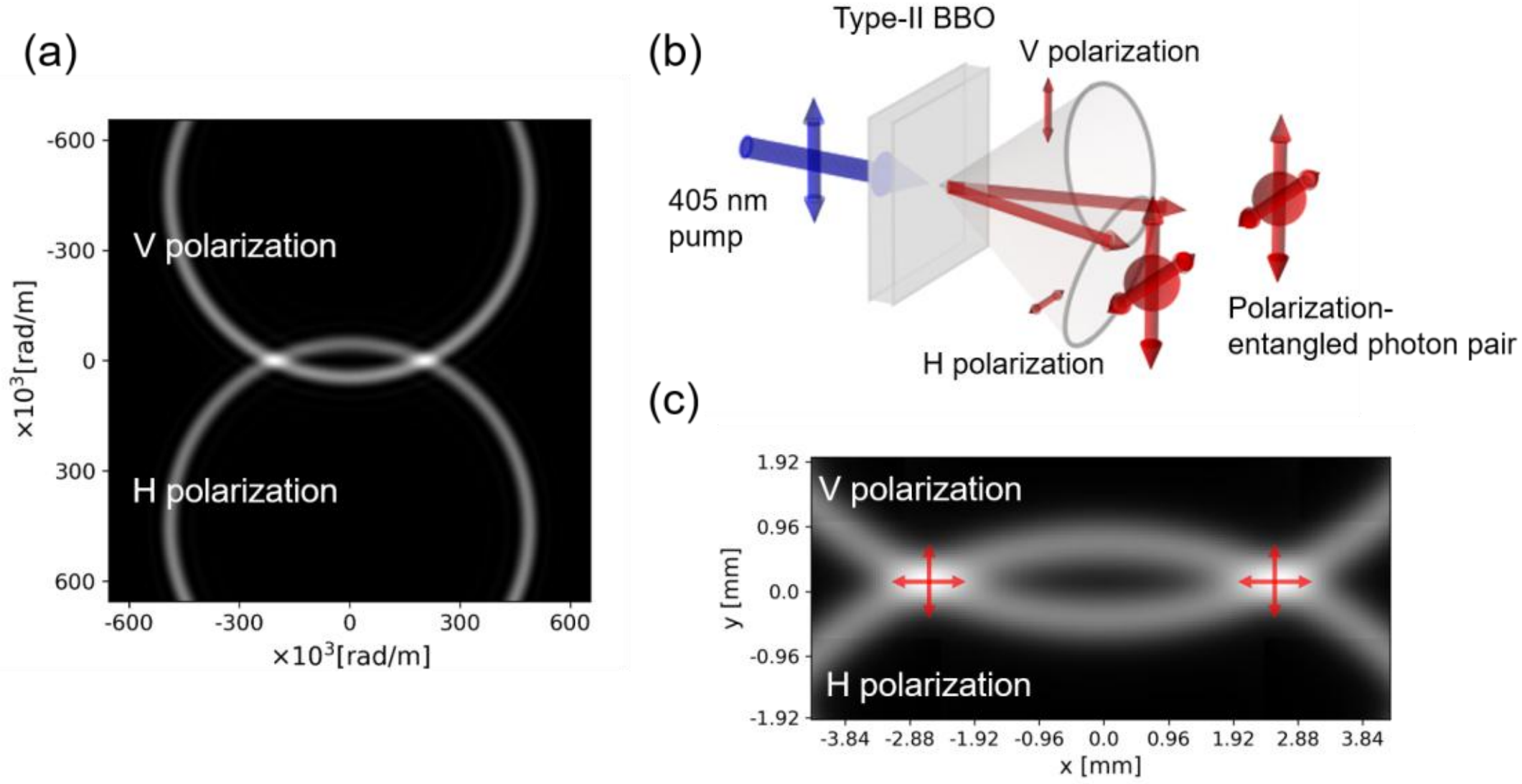

Fig. 1. (a) Schematic illustration of Type-II SPDC in a BBO crystal. Polarization-entangled photon pairs are generated in the overlap regions (A and B) of the signal and idler emission cones. (b) Calculated momentum-space intensity distributions of the signal and idler photons. (c) Calculated real-space intensity distributions evaluated at an observation plane located 10 cm from the crystal output surface. The calculations were performed for an optic-axis angle of $\theta = 41.8°$, a pump wavelength of $\lambda_p = 405$ nm, signal and idler wavelengths of $\lambda_{s,i} = 810$ nm, and a crystal length of $L = 2$ mm.

In Type-II SPDC, the signal and idler photons are emitted symmetrically with respect to the pump axis owing to momentum conservation. In this study, we define the photon positions emitted into the positive- and negative- $x$ directions as $\boldsymbol{\rho}_A = (x_A, y_A)$ and $\boldsymbol{\rho}_B = (x_B, y_B)$, respectively. Since the signal (H-polarized) and idler (V-polarized) photons are emitted into overlapping spatial modes, the biphoton state at positions $(\boldsymbol{\rho}_A, \boldsymbol{\rho}_B)$ is described by a coherent superposition of two indistinguishable SPDC emission processes,

$$|\Psi(\boldsymbol{\rho}_A, \boldsymbol{\rho}_B)\rangle = \psi_p(\boldsymbol{\rho}_A, \boldsymbol{\rho}_B)|\boldsymbol{\rho}_A\rangle_H|\boldsymbol{\rho}_B\rangle_V + \psi_p(\boldsymbol{\rho}_B, \boldsymbol{\rho}_A)|\boldsymbol{\rho}_B\rangle_H|\boldsymbol{\rho}_A\rangle_V. \quad (3)$$

Here, $\psi_p(\boldsymbol{\rho}_A, \boldsymbol{\rho}_B)$ represents the probability amplitude for detecting the H-polarized photon at $\boldsymbol{\rho}_A$ and the V-polarized photon at $\boldsymbol{\rho}_B$, whereas $\psi_p(\boldsymbol{\rho}_B, \boldsymbol{\rho}_A)$ represents the exchanged process. The amplitude-balance and relative phase between these two emission processes are key quantities that determine the polarization-entanglement purity. We therefore introduce

$$C(\boldsymbol{\rho}_A, \boldsymbol{\rho}_B) = \frac{|\psi_\rho(\boldsymbol{\rho}_A, \boldsymbol{\rho}_B)|^2 - |\psi_\rho(\boldsymbol{\rho}_B, \boldsymbol{\rho}_A)|^2}{|\psi_\rho(\boldsymbol{\rho}_A, \boldsymbol{\rho}_B)|^2 + |\psi_\rho(\boldsymbol{\rho}_B, \boldsymbol{\rho}_A)|^2}, \tag{4}$$

and

$$\phi(\boldsymbol{\rho}_A, \boldsymbol{\rho}_B) = \arg[\psi_\rho(\boldsymbol{\rho}_A, \boldsymbol{\rho}_B)] - \arg[\psi_\rho(\boldsymbol{\rho}_B, \boldsymbol{\rho}_A)], \tag{5}$$

where $C(\boldsymbol{\rho}_A, \boldsymbol{\rho}_B)$ characterizes the amplitude-balance between the two emission processes and $\phi(\boldsymbol{\rho}_A, \boldsymbol{\rho}_B)$ represents their relative phase difference. For a maximally polarization-entangled state, the two processes should contribute with equal amplitudes and a spatially uniform relative phase. Therefore, high polarization-entanglement purity is expected when $C(\boldsymbol{\rho}_A, \boldsymbol{\rho}_B) \approx 0$ and $\phi(\boldsymbol{\rho}_A, \boldsymbol{\rho}_B)$ remains nearly constant over the collected spatial region. In the following, we investigate the spatial distributions of $C(\boldsymbol{\rho}_A, \boldsymbol{\rho}_B)$ and $\phi(\boldsymbol{\rho}_A, \boldsymbol{\rho}_B)$ to identify the regions suitable for high-purity polarization entanglement.

Figures 2(a1)–(a3) show the spatial dependence on $\boldsymbol{\rho}_B = (x_B, y_B)$ for a fixed photon position $\boldsymbol{\rho}_A = \boldsymbol{\rho}_{A0}$, where $\boldsymbol{\rho}_{A0}$ is chosen at the point of maximum intensity in the overlap region of the two SPDC rings. Figure 2(a2) shows the spatial distribution of the $C(\boldsymbol{\rho}_{A0}, \boldsymbol{\rho}_B)$, whereas Fig. 2(a3) shows the corresponding $\phi(\boldsymbol{\rho}_A, \boldsymbol{\rho}_B)$. As shown in Fig. 2(a2), the $C(\boldsymbol{\rho}_{A0}, \boldsymbol{\rho}_B)$ becomes small in the cross-shaped region centered around $\boldsymbol{\rho}_B = (-x_{A0}, y_{A0})$, indicating a nearly equal contribution of the two emission processes. Figure 2(a3) shows that the $\phi(\boldsymbol{\rho}_A, \boldsymbol{\rho}_B)$ varies predominantly along the $y$-direction, whereas its variation along the $x$-direction is negligible. Therefore, for the fixed photon position $\boldsymbol{\rho}_{A0}$, high-purity polarization entanglement can be obtained only by collecting photon pairs from a horizontally elongated region centered at $\boldsymbol{\rho}_B = (-x_{A0}, y_{A0})$, where both amplitude-balance and phase-uniformity are simultaneously satisfied. Figures 2(b1)–(b3) show the spatial dependence on the vertical coordinates $(y_A, y_B)$ for $\boldsymbol{\rho}_A = (x_{A0}, y_A)$ and $\boldsymbol{\rho}_B = (x_{B0}, y_B)$, where $x_{B0} = -x_{A0}$. As shown in Figs. 2(b2) and 2(b3), $C(\boldsymbol{\rho}_A, \boldsymbol{\rho}_B)$ remains small throughout the $(y_A, y_B)$ region, while the phase $\phi(\boldsymbol{\rho}_A, \boldsymbol{\rho}_B)$ remains nearly constant along the positively correlated direction in the $(y_A, y_B)$ space and varies along the negatively correlated direction. This indicates that the amplitude-balance condition is broadly satisfied in the $(y_A, y_B)$ region, whereas phase uniformity is preserved only along the positively correlated direction i.e., where $y_A - y_B$ is constant. Figures 2(c1)–(c3) show the spatial dependence on the horizontal coordinates $(x_A, x_B)$ for $\boldsymbol{\rho}_A = (x_A, y_{A0})$ and $\boldsymbol{\rho}_B = (x_B, y_{B0})$, where $y_{A0} = y_{B0}$. As shown in Figs. 2(c2) and 2(c3), $C(\boldsymbol{\rho}_A, \boldsymbol{\rho}_B)$ remains small throughout the $(x_A, x_B)$ region, while $\phi(\boldsymbol{\rho}_A, \boldsymbol{\rho}_B)$ exhibits almost no variation. This indicates that the collection range in the $x$-direction is not significantly restricted, since both amplitude-balance and phase-uniformity are maintained throughout the $(x_A, x_B)$ region. Taken together, these results indicate that the amplitude-balance and phase-uniformity conditions restrict the collection region primarily in the $y$-direction, whereas the collection range in the $x$-direction is largely unrestricted. Consequently, a horizontally elongated slit aperture centered at the ring-overlap region is expected to provide both high photon-pair collection efficiency and high polarization-entanglement purity.

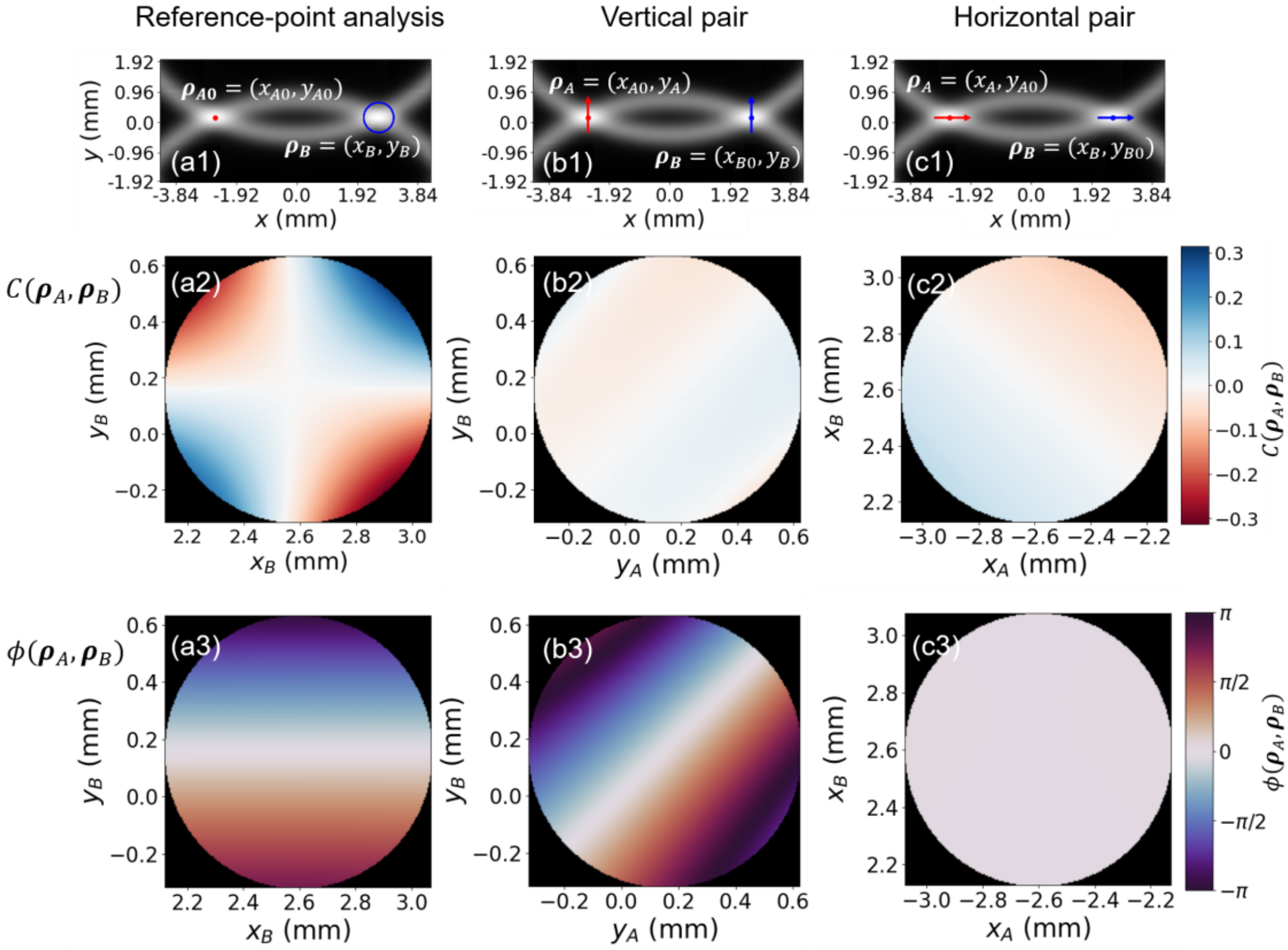

Fig. 2. Spatial distributions of the amplitude-balance parameter $C(\boldsymbol{\rho}_A, \boldsymbol{\rho}_B)$ and the relative phase $\phi(\boldsymbol{\rho}_A, \boldsymbol{\rho}_B)$ calculated from the real-space biphoton wavefunction. (a1) Analysis region $\boldsymbol{\rho}_B$ (blue) for a fixed photon position $\boldsymbol{\rho}_{A0}$ (red) located at the center of the overlap region of the two SPDC rings. (a2) Spatial distribution of $C(\boldsymbol{\rho}_{A0}, \boldsymbol{\rho}_B)$. (a3) Spatial distribution of $\phi(\boldsymbol{\rho}_{A0}, \boldsymbol{\rho}_B)$. (b1) Analysis region corresponding to $\boldsymbol{\rho}_A = (x_{A0}, y_A)$ and $\boldsymbol{\rho}_B = (x_{B0}, y_B)$, where $x_{B0} = -x_{A0}$. (b2) Spatial distribution of $C(x_{A0}, y_A, x_{B0}, y_B)$. (b3) Spatial distribution of $\phi(x_{A0}, y_A, x_{B0}, y_B)$ . (c1) Analysis region corresponding to $\boldsymbol{\rho}_A = (x_A, y_{A0})$ and $\boldsymbol{\rho}_B = (x_B, y_{B0})$, where $y_{A0} = y_{B0}$. (c2) Spatial distribution of $C(x_A, y_{A0}, x_B, y_{B0})$. (c3) Spatial distribution of $\phi(x_A, y_{A0}, x_B, y_{B0})$.

Based on the above discussion, the spatial distribution of the two-photon polarization state suggests that the collection region should be restricted in the $y$-direction while allowing a wider collection range in the $x$-direction. To verify this design principle, we theoretically compared the relationship between the $S$ parameter and the brightness for circular and $x$-elongated elliptical apertures. The aperture centers were set at the positions where the biphoton intensity was maximum. The $S$ parameter was calculated as the maximum value of the CHSH expression using the Horodecki criterion [28,32].

Figure 3(a) shows the calculated results for circular and elliptical apertures. The horizontal axis represents the brightness normalized to that obtained with the largest aperture, and the vertical axis represents the calculated $S$. The blue solid curve represents the results for circular apertures with different diameters, whereas the orange solid curve represents the results for elliptical apertures with a fixed length along the major axis ($x$-direction) and a variable length along the minor axis ($y$-direction). The red triangle and green circle in Fig. 3(a) correspond to the equal-area aperture configurations shown in Figs. 3(b) and 3(c), respectively. As the aperture size increases, the brightness increases whereas the $S$ decreases for both aperture geometries. This behavior arises because the larger collection region includes areas where the amplitude-balance between the two polarization components is disrupted and the phase-uniformity along the $y$-direction is degraded. Furthermore, in contrast to the circular aperture, an $x$-elongated elliptical aperture achieves a larger $S$ while maintaining comparable brightness. This improvement is attributed to the preferential collection of photon pairs from regions where both the amplitude-balance and phase-uniformity are preserved, as predicted by the real-space two-photon polarization-state distribution shown in Fig. 2. These results confirm that aperture geometry designed on the basis of the real-space two-photon polarization-state distribution can improve the trade-off between entanglement purity and photon-pair collection efficiency.

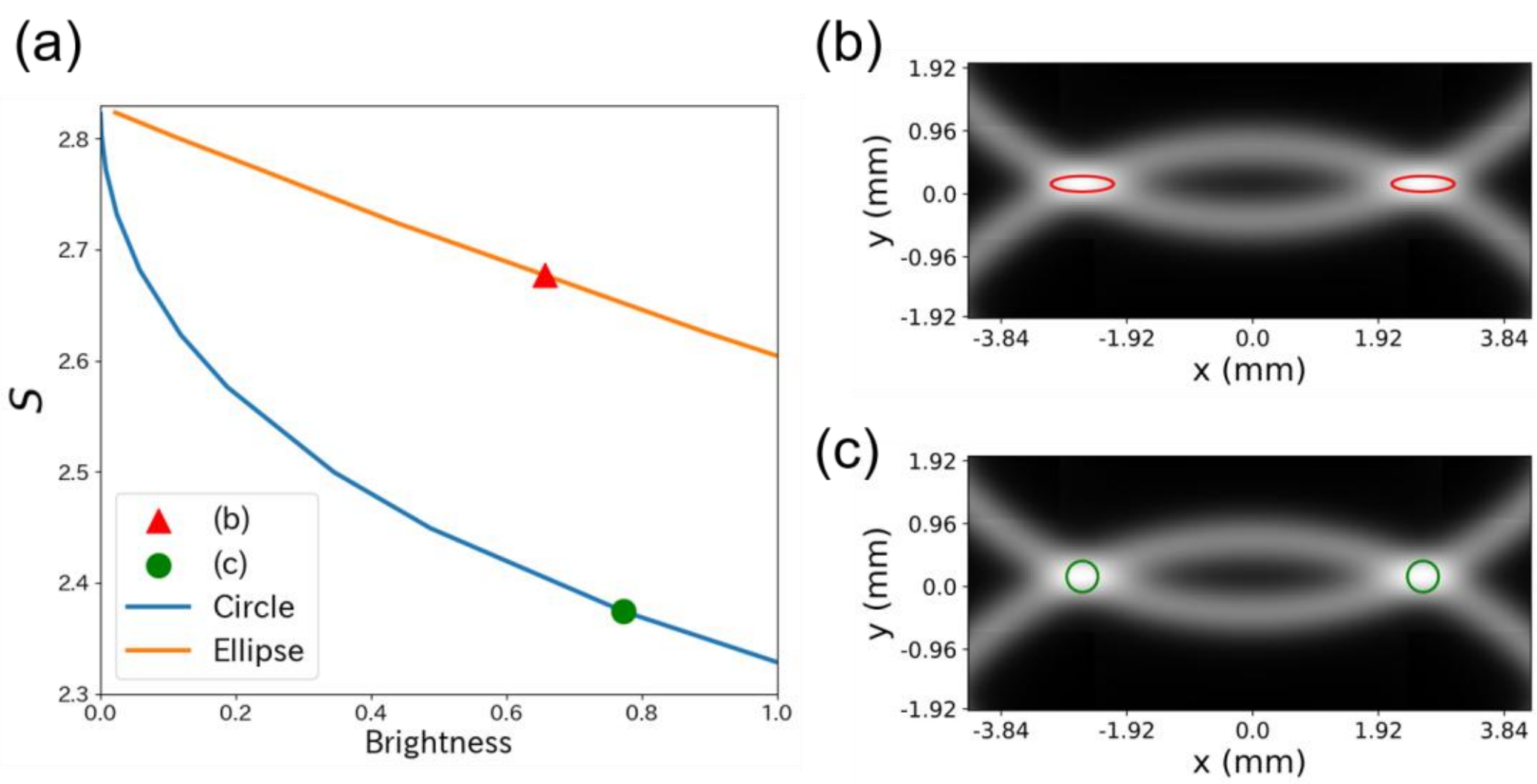

Fig. 3. Relationship between brightness and the $S$ parameter for different real-space aperture geometries. (a) Calculated $S$ for circular and elliptical apertures. The horizontal axis represents the brightness normalized to the value obtained with the largest circular aperture, whereas the vertical axis represents the calculated $S$. The blue curve corresponds to varying the diameter of the circular aperture, while the orange curve corresponds to varying the $y$-axis length of the elliptical aperture with the $x$-axis length fixed. The red triangle and green circle indicate the aperture configurations shown in (b) and (c), respectively.

## 3. Experimental setup

To examine whether the aperture design derived from the amplitude and phase structures of the real-space two-photon polarization-state distribution can improve polarization-entanglement purity, we experimentally investigated the brightness dependence of the CHSH Bell parameter $S$ for two aperture geometries: a circular aperture (iris) and a slit that restricts the spatial mode in the $y$-direction while allowing a wider collection range in the $x$-direction. In the experiment, a slit was used as a practical implementation of the $x$-elongated aperture geometry predicted by the theoretical analysis.

Figure 4 shows the experimental setup. Photon pairs were detected using two Si avalanche photodiodes (Si-APDs, SPCM-AQRH, Excelitas), and coincidence counts between the two detectors were analyzed using a Time Tagger 20 (Swabian Instruments). The coincidence time window was set to 3 ns. A band-pass filter (BPF, FBH810-10, Thorlabs) was used to select photon pairs centered at 810 nm. A 2-mm-thick Type-II BBO crystal (BBO 1) with an optic-axis angle of 41.8° was used as the SPDC source. To compensate for the temporal walk-off between the orthogonally polarized photons, two 1-mm-thick BBO crystals (BBO 2 and BBO 3) were inserted into the signal and idler paths. A 6-mm-diameter iris was placed in front of the coupling lens to suppress photons entering from unwanted optical paths. For the aperture measurements, the diameter of the iris was varied between 3 and 4.5 mm (3, 3.5, 4, and 4.5 mm), whereas the slit width was varied between 1.75 and 4 mm (1.75, 2.25, 3, and 4 mm). Two polarization measurement settings were chosen for each of the signal and idler photons, and the CHSH parameter was evaluated from the measured coincidence counts [18,29,33]. The projection angles for the polarization measurements at each detector were set to $\theta_{A1} = 22.5°$, $\theta_{A2} = 67.5°$, $\theta_{B1} = 0°$, and $\theta_{B2} = 45°$.

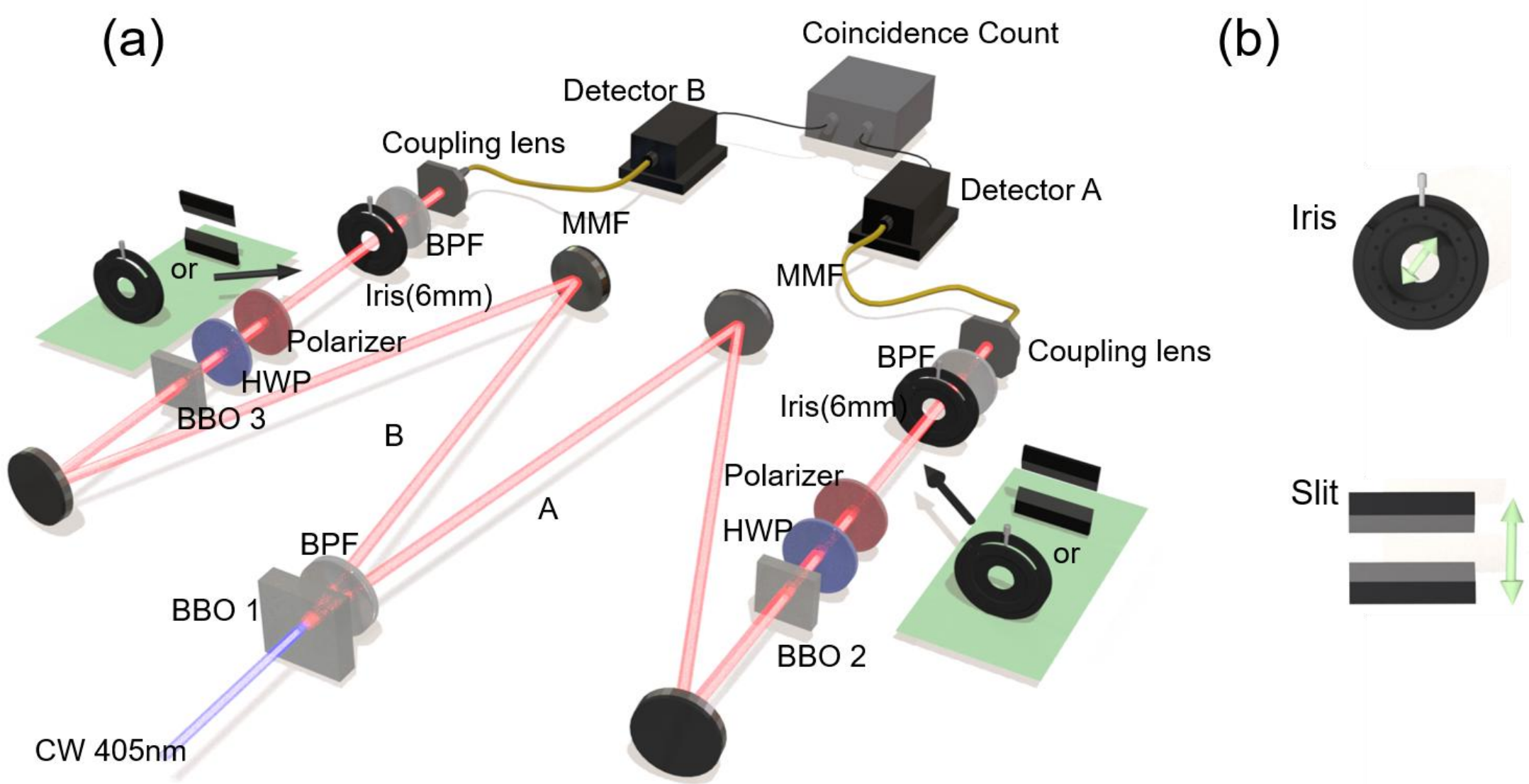

Fig. 4. Experimental setup and aperture geometries used to investigate the effect of aperture geometry on polarization entanglement. (a) Experimental setup. Photon pairs generated by Type-II SPDC are transmitted through either a circular aperture (iris) or a slit aperture before being coupled into multimode fibers and detected by Si avalanche photodiodes. The Bell parameter is evaluated from coincidence counts between detectors A and B. A band-pass filter (BPF, 810 nm center wavelength, 10 nm bandwidth) is used to select photon pairs centered at 810 nm. (b) Circular iris and slit apertures used for the measurements.

## 4. Results and Discussion

Figure 5 shows the experimental results. The brightness was defined as the coincidence count rate normalized by that obtained with the largest iris aperture. When a slit was used to restrict the collection region in one direction, a larger $S$ parameter was obtained than that achieved with an iris at a comparable brightness. As the slit width increased, however, the spatial selectivity became limited by the iris or collimating lens upstream of the BPF. As a result, the $S$ parameter decreased, and the measured values approached those obtained with an iris.

Spatial filtering in the $y$-direction with a slit selectively extracts spatial regions where the two-photon polarization state is nearly uniform. In contrast, a circular aperture collects photons from a wider spatial region that includes larger phase variations, resulting in a reduction of the polarization-entanglement purity. This tendency agrees with the theoretical prediction shown in Fig. 3, where a horizontally elongated aperture was found to be more effective than a circular aperture in preserving entanglement purity.

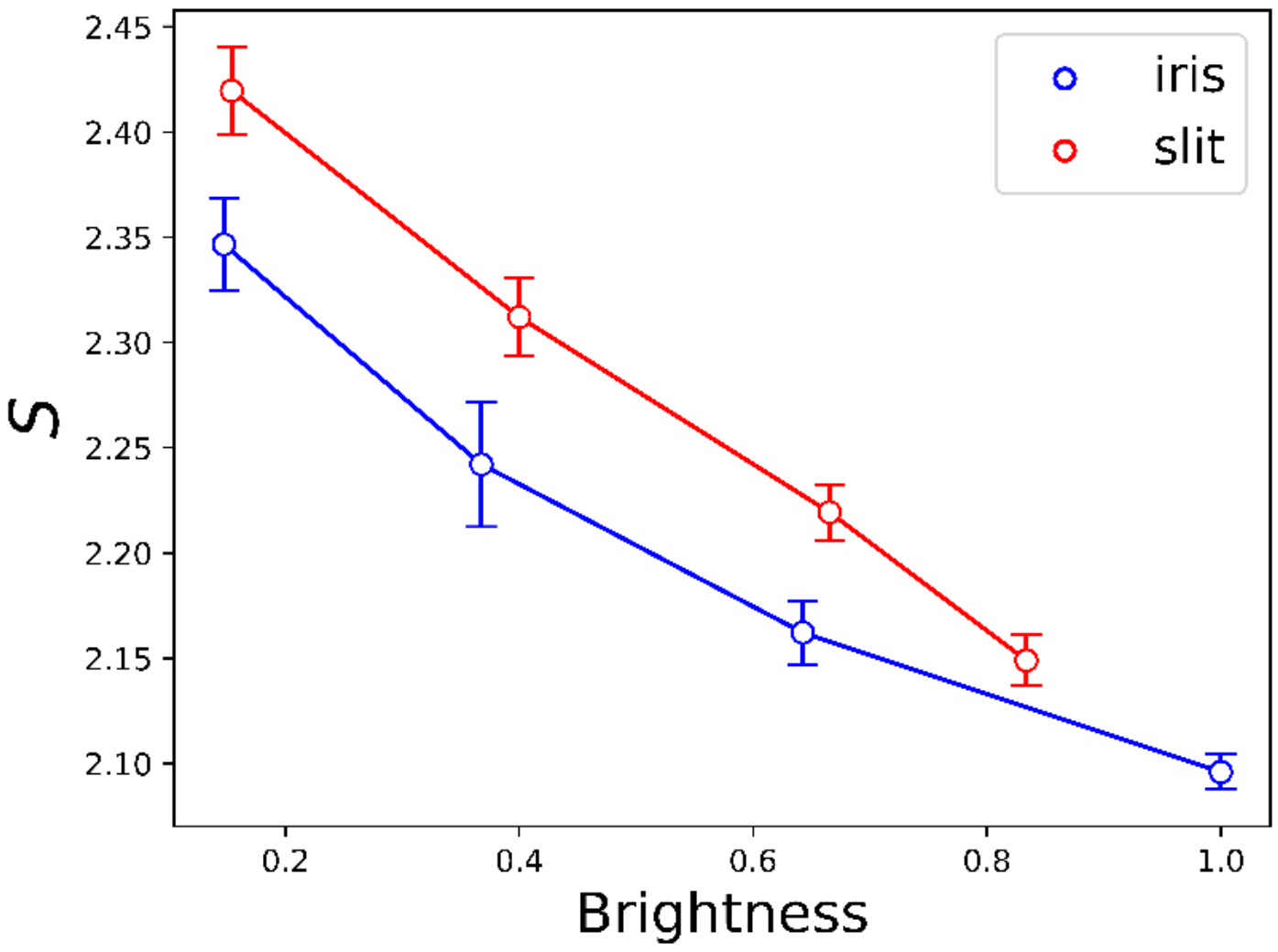

Fig. 5. Experimental dependence of the CHSH Bell parameter $S$ on brightness for iris and slit apertures. Blue circles represent the results obtained using circular apertures (iris diameters of 3, 3.5, 4, and 4.5 mm), whereas red circles represent the results obtained using slit apertures (slit widths of 1.75, 2.25, 3, and 4 mm). The brightness is defined as the coincidence count rate normalized to that obtained with the largest iris aperture. Error bars indicate the standard deviation of five repeated measurements. At a given brightness, the slit aperture yields a larger $S$ parameter than the iris aperture.

The measured absolute $S$, as well as the enhancement ratio in $S$ obtained with the slit aperture relative to the iris aperture, was smaller than predicted by the theoretical model (Fig. 3). One possible reason for this discrepancy is that the theoretical model assumes monochromatic pump, signal, and idler photons, whereas the experimental setup cannot completely isolate a single-frequency component. In Type-II SPDC using a BBO crystal, the joint spectral amplitude of the signal and idler photons becomes asymmetric with respect to the correlation axis due to birefringence, resulting in incomplete overlap of the spectral wavefunctions associated with the two SPDC emission processes [34–36]. Furthermore, because the phase-matching condition varies with frequency, correlations arise between the polarization and frequency degrees of freedom. Consequently, the residual correlations among the frequency, polarization, and spatial degrees of freedom may lead to a mixed polarization state and may reduce both $S$ and the improvement achieved by the slit aperture [37]. Despite the quantitative discrepancy, the experimental results qualitatively agree with the theoretical prediction, confirming that the amplitude and phase structures of the real-space two-photon polarization-state distribution provide a practical guideline for aperture design to improve polarization-entanglement purity.

## 5. Conclusion

In this study, we calculated the real-space biphoton wavefunction in SPDC and determined the real-space distribution of the two-photon polarization state. Based on this distribution, we theoretically and experimentally demonstrated that restricting the collection region along the direction of strong polarization–position correlation using an appropriately designed aperture enables the simultaneous achievement of high polarization-entanglement purity and high photon-pair collection efficiency. These findings provide practical design guidelines for aperture geometries for the efficient generation of high-quality polarization-entangled photon pairs. Furthermore, the present framework may provide a basis for the design and control of spatially structured two-photon polarization states, with potential applications to the generation, manipulation, transfer, and imaging of structured photonic quantum states [9, 38-41].

**Funding.** Japan Society for the Promotion of Science (22H01981, 25H01608, 25K01687, 26K22732); JKA Foundation (2024M-374); Toshiba Devices & Storage Academic Incentive System 2024-2026.

**Acknowledgements.** The authors thank E. Abe for helpful discussions on single-photon detection. The authors also thank M. Komine and N. Sugi for their contributions to the initial development of the experiment setup.

**Disclosures.** The authors declare no conflicts of interest.